\documentclass[preprint]{aastex}

\begin{document}
\title{Gas Gaps in the Protoplanetary Disk around the Young Protostar HL Tau}

\author{Hsi-Wei Yen\altaffilmark{1}, Hauyu Baobab Liu\altaffilmark{2}, Pin-Gao Gu\altaffilmark{1}, Naomi Hirano\altaffilmark{1}, Chin-Fei Lee\altaffilmark{1}, Evaria Puspitaningrum\altaffilmark{3}, Shigehisa Takakuwa\altaffilmark{1,4}}

\altaffiltext{1}{Academia Sinica Institute of Astronomy and Astrophysics, P.O. Box 23-141, Taipei 10617, Taiwan; hwyen@asiaa.sinica.edu.tw} 
\altaffiltext{2}{European Southern Observatory (ESO), Karl-Schwarzschild-Str. 2, D-85748 Garching, Germany}
\altaffiltext{3}{Department of Astronomy, Faculty of Mathematics and Natural Sciences, Institut Teknologi Bandung, Jl. Ganesha 10, Bandung 40132 Indonesia}
\altaffiltext{4}{Department of Physics and Astronomy, Graduate School of Science and Engineering, Kagoshima University, 1-21-35 Korimoto, Kagoshima, Kagoshima 890-0065, Japan}


\begin{abstract}
We have analyzed the HCO$^+$ (1--0) data of the Class I--II protostar, HL Tau, obtained from the Atacama Large Millimeter/Submillimeter Array long baseline campaign. 
We generated the HCO$^+$ image cube at an angular resolution of $\sim$0\farcs07 ($\sim$10 AU), 
and performed azimuthal averaging on the image cube 
to enhance the signal-to-noise ratio and measure the radial profile of the HCO$^+$ integrated intensity. 
Two gaps at radii of $\sim$28 AU and $\sim$69 AU and a central cavity are identified in the radial intensity profile.  
The inner HCO$^+$ gap is coincident with the millimeter continuum gap at a radius of 32 AU.  
The outer HCO$^+$ gap is located at the millimeter continuum bright ring at a radius of 69 AU and overlaps with the two millimeter continuum gaps at radii of 64 AU and 74 AU.
On the contrary, 
the presence of the central cavity is likely due to the high optical depth of the 3 mm continuum emission and not the depletion of the HCO$^+$ gas. 
We derived the HCO$^+$ column density profile from its intensity profile.
From the column density profile, the full-width-half-maximum widths of the inner and outer HCO$^+$ gaps are both estimated to be $\sim$14 AU, 
and their depths are estimated to be $\sim$2.4 and $\sim$5.0.
These results are consistent with the expectation from the gaps opened by forming (sub-)Jovian mass planets, 
while placing tight constraints on the theoretical models solely incorporating the variation of dust properties and grain sizes. 
\end{abstract}

\section{Introduction}
HL Tau is a Class I--II protostellar source located in the northern part of the L1551 region in the Taurus molecular cloud at a distance of $\sim$140 pc \citep{Hayashi09}.
It is still embedded in a protostellar envelope with a size of $\sim$3000 AU and surrounded by a circumstellar disk with a size of $\sim$100 AU \citep{Mundy96, Wilner96, Carrasco09, Kwon11}. 
Previous interferometric observations have found infalling motion in the protostellar envelope \citep{Hayashi93}. 
In addition, 
jets and bipolar outflows launched from this protostellar source have also been observed at optical and infrared wavelengths and in molecular lines \citep{Monin96, Takami07, Hayashi09, Lumbreras14}.
The presence of the infall and the outflow implies that HL Tau is still in the active mass accretion phase.

Recent observations of the millimeter and submillimeter continuum emission with the Atacama Large Millimeter/submillimeter Array (ALMA) at high angular resolutions of $\sim$0\farcs03--0\farcs07, corresponding to a physical size of $\sim$5--10 AU, discovered a series of ringlike gaps in the disk around HL Tau \citep{ALMA15, Akiyama15}. 
The presence of the gaps suggests the depletion of material or the variation of dust properties at these radii.
It is considered as hints of forming planets in the disk. 
Therefore, HL Tau becomes a promising candidate of ongoing planet formation. 

Subsequent theoretical studies to understand the origin of the ringlike structures predict the presence of multiple, sub-Jovian mass ($M_J$) planets in the disk \citep{Dipierro15, Kanagawa15, Tamayo15, Jin16}.  
The other studies claim that gravitational instabilities \citep{Takahashi14}, pebble growths \citep{Zhang15}, or dust aggregates \citep{Okuzumi15} can explain the observed features. 
A direct search for planetary bodies in the gaps in the disk around HL Tau has been performed with imaging observations in $L^\prime$ band. 
No planet was found, and the observations put the upper limits on the planet mass to be 10--15 $M_J$ \citep{Testi15}. 

However, the information on the gas component, the major mass content in the disk, in these gaps of the millimeter continuum emission remains unclear. 
Even though the HCO$^+$ (1--0) emission, tracing the gas component in the disk around HL Tau, is also detected at standard-of-rest velocities ($V_{\rm LSR}$) of 2--12 km s$^{-1}$ in the ALMA observations, 
its image was generated at a lower angular resolution of $\sim$0\farcs25, corresponding to 35 AU, because of the limited signal-to-noise (S/N) ratio \citep{ALMA15}.
This HCO$^+$ (1--0) image indeed clearly shows the Keplerian rotation of the disk \citep{ALMA15, Pinte15}, 
but its resolution is too coarse to identify the counterparts of these continuum gaps in the image. 
In this letter, we report the radial intensity profiles and reconstructed intensity map of the HCO$^+$ (1--0) emission in the disk around HL Tau at a high angular resolution of $\sim$0\farcs07, revealing the counterparts of the continuum gaps in the gas component.

\section{Observations}
The HL Tau data analyzed here were retrieved from the public data release of the ALMA Science Verification observations (project code: 2011.0.00015.SV). 
The details of the observations have been described by \citet{ALMA15}.  
We retrieved the 3 mm continuum image directly from the ALMA data archive, 
while re-generated the HCO$^+$ (1--0; 89.188526 GHz) and CO (1--0; 115.271202 GHz) image cubes at higher angular resolutions than those presented by \citet{ALMA15}.  
Our HCO$^+$ and CO image cubes were generated by Fourier-transforming their calibrated visibility data at a velocity resolution of 0.4 km s$^{-1}$ with the Briggs robustness parameter of $-1$ and without tapering, using the tasks in Common Astronomy Software Applications (CASA). 
We did not perform self-calibration on their visibility data because self-calibration did not improve their S/N ratios, as demonstrated by \citet{ALMA15}.
Because the CO image cube was used to derive the temperature profile, 
we did not perform the continuum subtraction on its visibility data, as explained below, 
while the HCO$^+$ data were continuum subtracted using the CASA task {\it uvcontsub}.
Because there is no clear detection above S/N ratios of 3 in the image cubes, no ``CLEAN'' was performed. 
The angular resolutions of the HCO$^+$ and CO images are $\sim$0\farcs1 $\times$ 0\farcs05 and $\sim$0\farcs07 $\times$ 0\farcs04, respectively, 
and their noise levels are $\sim$2 and $\sim$6 mJy Beam$^{-1}$ per channel. 

\section{Results}
\subsection{HCO$^+$ (1--0) Intensity Profile}
\begin{figure}
\epsscale{0.8}
\plotone{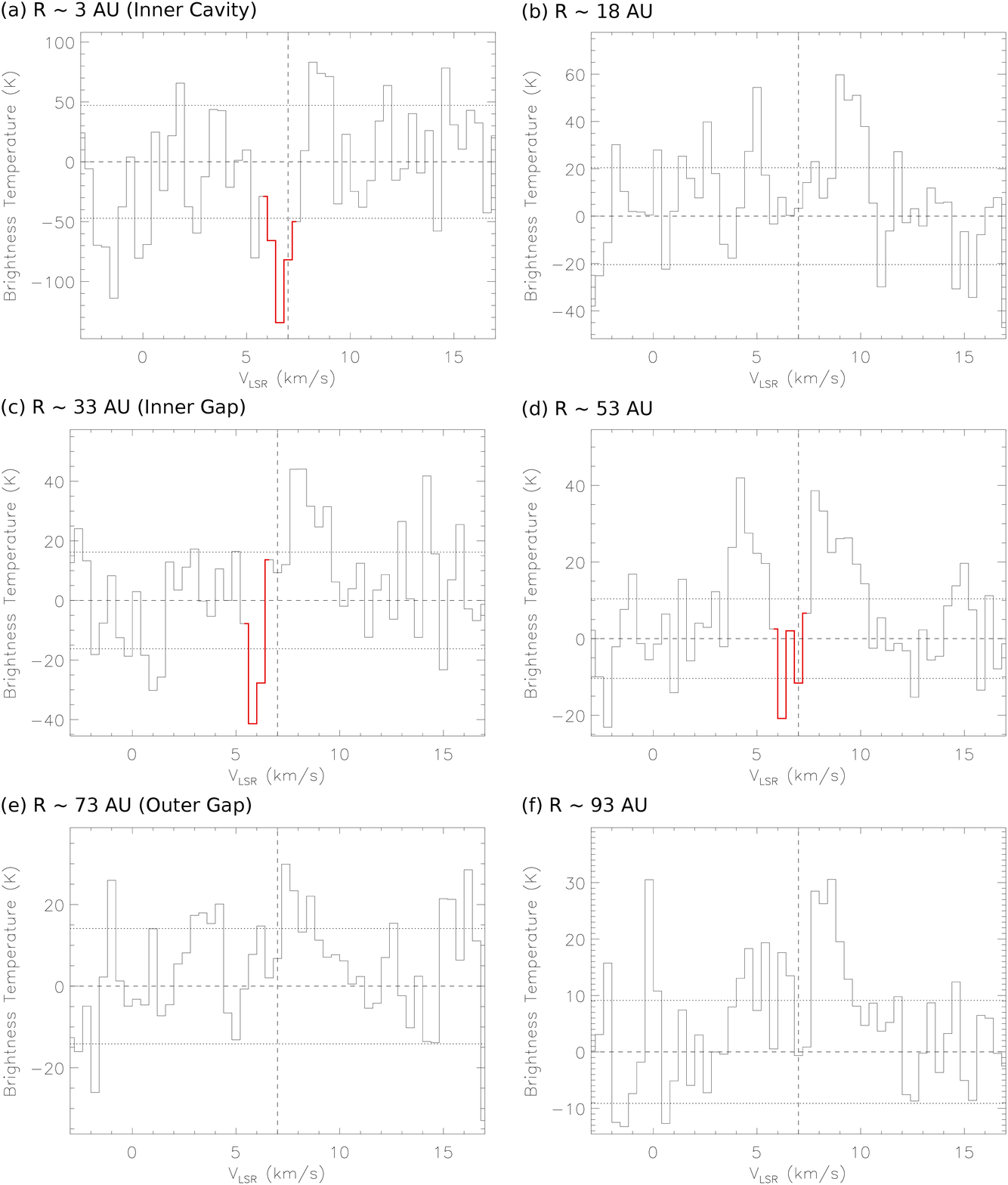}
\caption{Azimuthally-averaged spectra of the HCO$^+$ (1--0) emission in the disk around HL Tau observed with ALMA. The central radius of the radial bin of the each spectrum is labeled above each panel. (a), (c), and (e) present the spectra in the central cavity and the two gaps. (b), (d), and (f) present the spectra in the bright regions near the gaps. Vertical and horizontal dashed lines denote the zero intensity and the systemic velocity of HL Tau ($V_{\rm LSR}$ of 7 km s$^{-1}$), respectively. Two horizontal dotted lines denote $\pm$1$\sigma$ noise levels of the individual spectra. Red histograms present the channels which show absorption and were excluded when we measured the integrated intensity.}
\label{spec}
\end{figure}

Because there is no clear detection at a S/N ratio $>$ 3 in our high-resolution HCO$^+$ image cube, 
we performed azimuthal averaging on the image cube to enhance the S/N ratio and to measure its radial intensity profile. 
The disk around HL Tau is inclined by 47$\arcdeg$ from the plane of the sky, 
and its major axis has a position angle of 138$\arcdeg$ from north to east \citep{ALMA15}. 
These angles were adopted to correct the projection of the disk on the plane of the sky, 
and the de-projected radius of each pixel was computed. 
The size of each radial bin for azimuthal averaging was adopted to be the half of the beam size ($\sim$0\farcs035 or $\sim$5 AU).  
To minimize the confusion from the foreground or background molecular clouds and the associated outflows \citep{ALMA15}, 
only the pixels within $\pm$45$^\circ$ of the major axis were included in azimuthal averaging, where the velocities are blueshifted and redshifted away from the systemic velocity because of the rotation of the disk. 
Then we obtained the azimuthally averaged HCO$^+$ spectrum at each radial bin. 
The averaged spectra at representative radii are presented in Figure \ref{spec}.
The spectra at several radii show a negative dip centered at $V_{\rm LSR}$ $\sim$ 6--7 km s$^{-1}$, 
which has also been seen in the spectrum at the low angular resolution \citep{ALMA15}.
That can be due to absorption by foreground clouds or the outflow associated with HL Tau. 
We integrated the spectrum over $V_{\rm LSR}$ of 2--12 km s$^{-1}$ to measure the integrated intensity at each radial bin,  
and the velocity channels showing absorption (red histograms in Figure \ref{spec}) and those having values within the $\pm$1$\sigma$ noise level were excluded when we measured the integrated intensity.
The 1$\sigma$ noise level of each averaged spectrum was measured using the line-free channels in that spectrum.
The measured radial profile of the integrated intensity of the HCO$^+$ emission is presented in Figure \ref{Ir}a.
The uncertainty of the integrated intensity at each radial bin was estimated as $\sqrt{N_{\rm int}} \times \sigma \times dv$, where $N_{\rm int}$ is the number of channels integrated, $\sigma$ is the noise level of the averaged spectrum, and $dv$ is the velocity width of one channel.

\begin{figure}
\epsscale{1}
\plotone{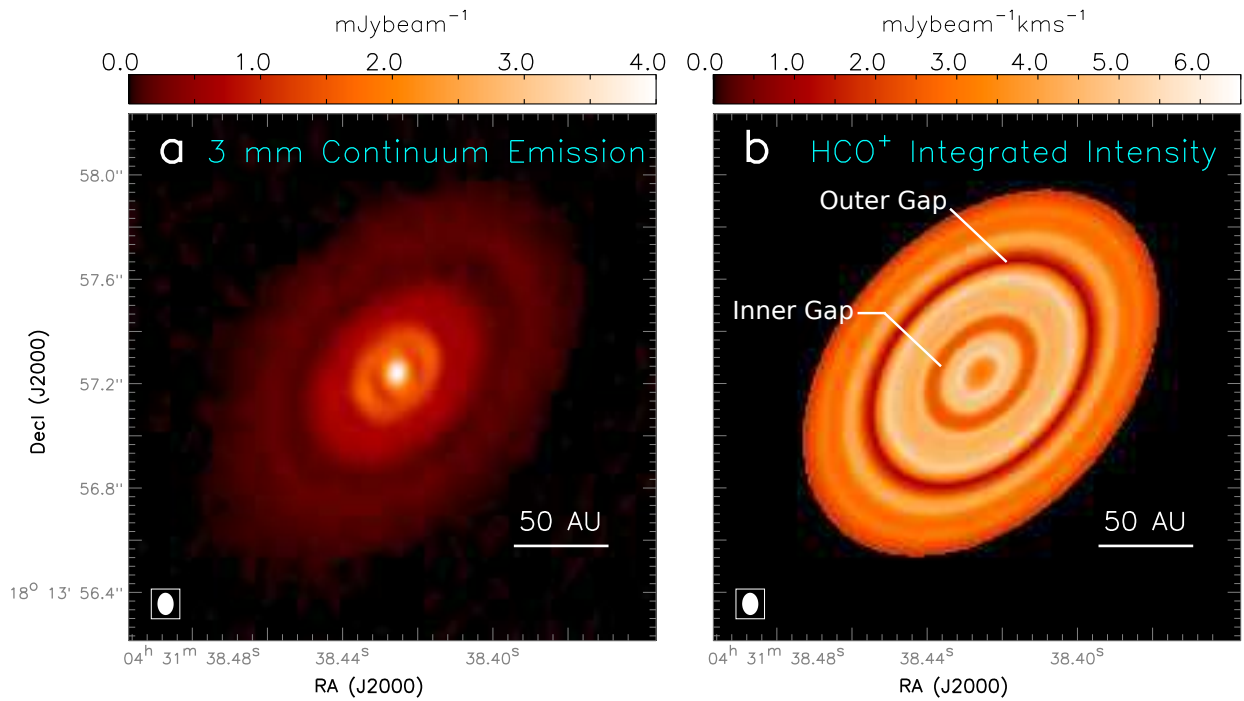}
\caption{(a) 3 mm continuum image and (b) our reconstructed high-resolution image of the HCO$^+$ (1--0) integrated intensity of the disk around HL Tau observed with ALMA at angular resolutions of $\sim$0\farcs07, corresponding to a physical scale of $\sim$10 AU.}
\label{map}
\end{figure}

For visualization, we reconstructed the integrated intensity map of the HCO$^+$ emission from its radial intensity profile, assuming the intensity distribution is axisymmetric. 
We first computed the de-projected radius of each pixel in the map, and the pixel value was assigned by linear interpolation between the two closest radial bins of the intensity profile.
We note that in the low-resolution HCO$^+$ image \citep{ALMA15}, the distribution of the HCO$^+$ emission is not fully symmetric, 
and the emission on the northwestern side is brighter than that on the southeastern side.  
Due to the limited S/N ratio in our high-resolution image cube, the asymmetry of the intensity distribution cannot be further investigated. 
Nevertheless, the hints of the gaps at $\sim$30 AU and $\sim$70 AU are seen on the both sides of the disk when we measured the HCO$^+$ intensity profile.
The reconstructed integrated intensity map of the HCO$^+$ emission is presented in Figure \ref{map} in comparison with the 3 mm continuum image of the HL Tau disk, which is obtained with the same ALMA observations and has a comparable angular resolution \citep{ALMA15}.
Our HCO$^+$ intensity map clearly shows two gaps and a cavity at the center. 

In our HCO$^+$ results, there is no counterpart of the innermost gap at a radius of $\sim$13 AU and the central emission peak seen in the 3 mm continuum image \citep{ALMA15, Akiyama15}.
Instead, the HCO$^+$ intensity map displays a $\sim$10 AU wide cavity at the center.
Likely, this is at least partially related to the high optical depth of the 3 mm dust continuum emission toward the center. 
The continuum-subtracted spectral-line image is essentially biased by features which are anti-correlated with the intensity distribution of the continuum emission.
Because of the increasing continuum intensity toward the center, 
we are expecting to see the following effects on the continuum-subtracted HCO$^+$ data, 
(1) absorption features at the peak of the continuum emission, 
(2) a suppressed contrast between the rings and gaps in the HCO$^+$ intensity, 
and (3) a flattening in the HCO$^+$ intensity profile towards the continuum peak (Figure \ref{Ir}a).
Indeed, the derived radial profile of the HCO$^+$ column density, as shown below, is steeper than its intensity profile (Figure \ref{Ir}), although the overall HCO$^+$ column density profile remains biased by the not as well constrained radial temperature profiles of the dust and the gas.
 
\subsection{HCO$^+$ Column Density Profile}
To estimate the column density of the HCO$^+$ gas, 
we assume that the dust grains and the gas are thermally well coupled and have the same temperature, and that the HCO$^+$ gas is thermalized for simplicity. 
The temperature profile ($T$) is adopted to be the radial profile of the brightness temperature of the CO emission without continuum subtraction, on the assumption that the CO + 3 mm continuum emission is optically thick and traces the temperature structure of the disk. 
The radial profiles of the CO brightness temperature without continuum subtraction was measured with the same method as that for the HCO$^+$ intensity profile, 
and is shown in Figure \ref{Tr}b. 
Because the bright continuum emission can bias the continuum-subtracted line intensity, 
we first estimated the radial profile of the optical depth of the 3 mm continuum emission ($\tau_{\rm 3mm}$) from its radial intensity profile and the temperature profile using the radiative transfer equation, 
\begin{equation}\label{tau}
T_{\rm b, 3mm} = (T - T_{\rm bg}) \cdot (1 - e^{-\tau_{\rm 3mm}}),
\end{equation}
where $T_{\rm b, 3mm}$ is the brightness temperature of the 3 mm continuum emission and $T_{\rm bg}$ is the temperature of the cosmic microwave background radiation ($\sim$2.73 K). 
The derived radial profile of the optical depth of the 3 mm continuum emission is shown in Figure \ref{Tr}a,
and its 1$\sigma$ uncertainty was estimated by propagating the uncertainties in $T$ and $T_{\rm b,3mm}$ through Equation \ref{tau}.
The continuum-subtracted brightness temperature of the HCO$^+$ emission ($T_{\rm b,HCO^+}$) can be described as 
\begin{equation}
T_{\rm b,HCO^+} = T_{\rm b, 89GHz} - T_{\rm b,3mm}, 
\end{equation}
where $T_{\rm b, 89GHz}$ is the brightness temperature at the frequency of the HCO$^+$ line before the continuum subtraction, 
and 
\begin{equation}
T_{\rm b, 89GHz} = (T - T_{\rm bg}) \cdot (1 - e^{-\tau_{\rm tot}}), 
\end{equation}
where $\tau_{\rm tot}$ is the total optical depth contributed by the 3 mm continuum and the HCO$^+$ line ($\tau_{\rm HCO^+}$), i.e.,  $\tau_{\rm tot} = \tau_{\rm 3mm} + \tau_{\rm HCO^+}$. 
Therefore, we corrected the effect of the continuum optical depth on the observed HCO$^+$ brightness temperature as 
\begin{equation}\label{tauhcop}
T_{\rm b,HCO^+} = (T - T_{\rm bg}) \cdot e^{-\tau_{\rm 3mm}} \cdot (1 - e^{-\tau_{\rm HCO^+}}), 
\end{equation}
and estimated $\tau_{\rm HCO^+}$ at each radial bin. 
$T$ and $T_{\rm b,HCO^+}$ were adopted to be the mean brightness temperature of the azimuthally-averaged CO and HCO$^+$ spectra within $V_{\rm LSR}$ of 2--12 km s$^{-1}$. 
Then, following the derivation in \citet{Mangum15}, we estimated the HCO$^+$ column density as 
\begin{equation}\label{nhcop}
2.2 \times 10^7 \cdot T \cdot \frac{e^{4.28/T}}{1-e^{-4.28/T}} \frac{\tau_{\rm HCO^+}}{1 - e^{-\tau_{\rm HCO^+}}} \int \tau_{\rm HCO^+} dv\ {\rm cm^{-2}}. 
\end{equation}
The derived profile of the HCO$^+$ column density is presented in Figure \ref{Ir}c, 
and its 1$\sigma$ uncertainty was estimated by propagating the uncertainties in $T_{\rm b,HCO^+}$, $T$, and $\tau_{\rm 3mm}$ through Equation \ref{tauhcop} and \ref{nhcop}.
The column density profile presented here has radial bins with a size of $\sim$0\farcs035 ($\sim$5 AU), which is the half of the beam size, and is azimuthally averaged over $\pm$45$\arcdeg$ of the major axis.

We note that the mm-sized dust grains, which are located closer to the mid-plane, can have a lower temperature than the gas located closer to the disk surface.
If that is the case, the dust temperature is lower than the measured $T$, and then $\tau_{\rm 3mm}$ is underestimated. 
The derived HCO$^+$ column density becomes higher. 
In addition, the absorption in the HCO$^+$ spectra (Figure \ref{spec}) is expected to be more significant at the radii of the continuum bright rings than the continuum gaps, and hence, the HCO$^+$ column density at the radii of the bright rings can be underestimated more than that at the gaps. 
These effects can lead to larger depths of the HCO$^+$ gaps than the values estimated in this work.

\subsection{HCO$^+$ Gaps}
The overall column density profile cannot be well represented by a single power-law function (Fig.~\ref{Ir}c).
Therefore, we treated the inner and outer HCO$^+$ gaps separately to derive their widths and depths. 
For each gap, we first fitted a power-law function to the column densities only at the radial bins near the gap (red dashed and dotted lines in Figure \ref{Ir}c) to interpolate the column densities within the gap.
The fitted power-law function was subtracted from the column density profile at those radial bins including the gap.
Then we fitted each gap in the residual with a Gaussian function, 
and derived the full-width-half-maximum (FWHM) of each gap and a column density at the center of each gap.
The contrast in the column density inside and outside the gaps is estimated by comparing the column density at the center of the gaps and that of the fitted power-law function at the same radius (Figure \ref{Ir}c).
We repeated this fitting process for a 10,000 times, where each time the radii and densities of the data points are randomly changed within the radial bins and the 1$\sigma$ uncertainties, respectively. 
With this we obtained the probability distributions of the fitting parameters, 
and the 10,000 iterations is sufficient to converge. 
Then we fitted Gaussian functions to these probability distributions, 
and the centers and standard deviations of the fitted Gaussian functions were adopted as the best-fit values and the 1$\sigma$ uncertainties.
With these processes,
the HCO$^+$ column density in the inner and outer gaps are estimated to be $\sim$(5$\pm$1) $\times$ 10$^{15}$ cm$^{-2}$ and $\sim$(1.4$\pm$0.4) $\times$ 10$^{15}$ cm$^{-2}$. 
Both HCO$^+$ gaps have similar FWHM widths of $\sim$14$\pm$7 AU. 
Assuming the HCO$^+$ abundance does not vary across the gaps, 
the contrast between the gas density inside and outside the gaps is a factor of $\sim$2.4$\pm$0.5 and $\sim$5.0$\pm$1.2.
The detailed fitting results are listed in Table \ref{gap}.

\section{Discussion}
\begin{figure}
\epsscale{0.5}
\plotone{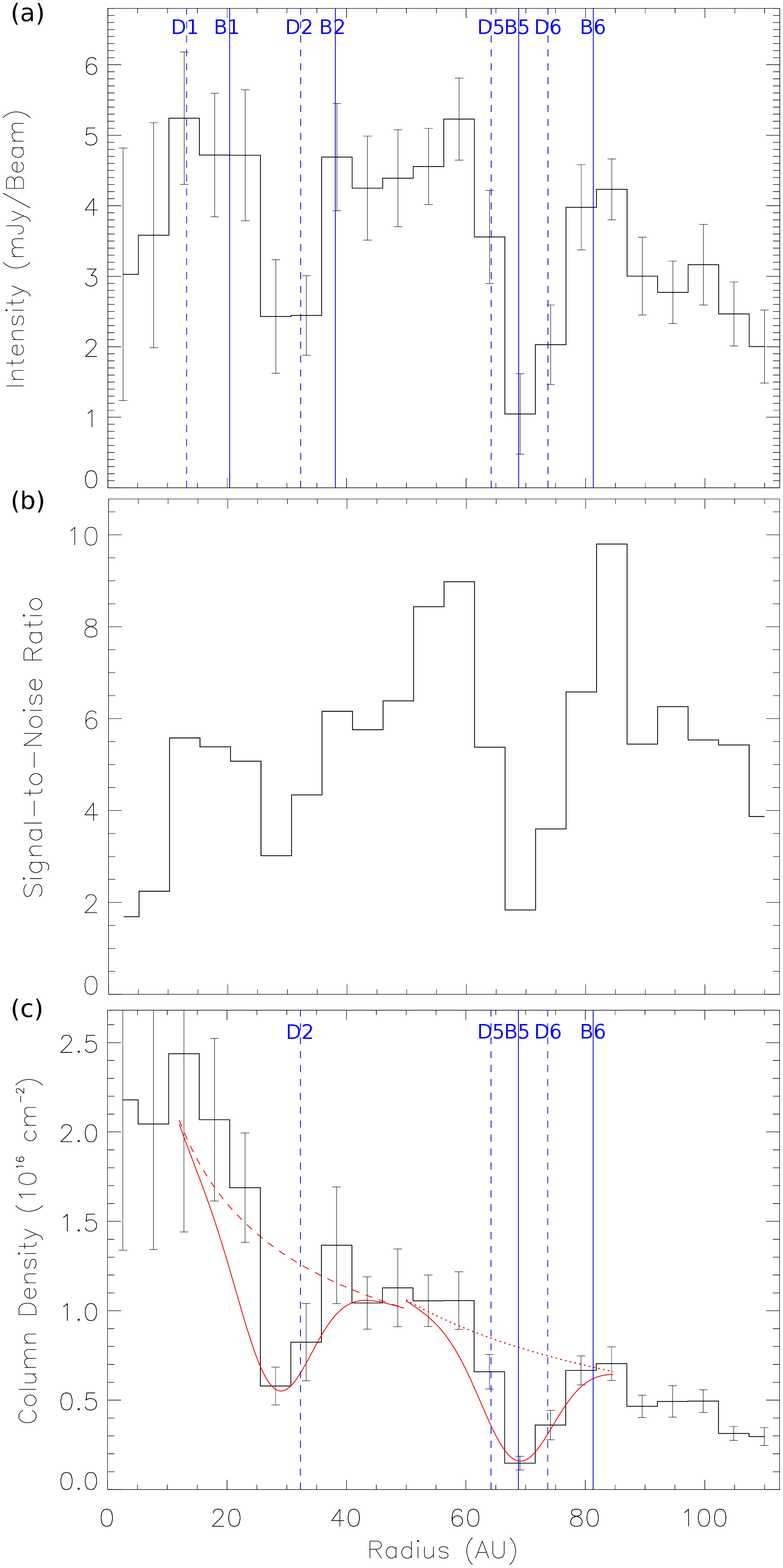} 
\caption{(a) Radial profile of the HCO$^+$ integrated intensity in the disk around HL Tau observed with ALMA. Vertical blue solid and dashed lines denote the radii of the bright rings and the gaps identified in the 1 mm continuum image observed with ALMA at an angular resolution of 0\farcs03 (corresponding to $\sim$5 AU; \citet{ALMA15}). (b) Radial profile of the S/N ratios of the HCO$^+$ integrated intensity. (c) Radial profile of the HCO$^+$ column density derived from the intensity profile. Red dashed and dotted lines present the fitted power-law density profiles outside the inner and outer gaps, respectively, and red solid lines delineate the observed gaps. The centers, widths, and depths of the gaps are listed in Table \ref{gap}. Error bars in panels (a) and (c) correspond to the 1$\sigma$ uncertainties.}
\label{Ir}
\end{figure}

The locations of two HCO$^+$ gaps are compared with those of the gaps and bright rings in the milimmeter continuum images (Figure \ref{Ir}). 
The inner HCO$^+$ gap is coincident with the continuum gap at a radius of 32.3 AU (D2 gap; \citet{ALMA15}). 
On the other hand, 
the outer HCO$^+$ gap is located at the bright continuum ring at a radius of 68.8 AU (B5 ring), 
and overlaps with the two continuum gaps at 64.2 AU and 73.7 AU (D5 and D6 gaps; \citet{ALMA15}). 
Our results indicate that the material, including both dust and gas, are depleted at the D2, D5, and D6 gaps.
By comparing with the gap depths estimated from the continuum data \citep{Kanagawa15, Pinte15}, 
the gas gaps are comparable to or shallower than the dust gaps. 

Hydrodynamical simulations and analytical frameworks have suggested that these gaps can be opened due to tidal interactions with sub-Jovian or Saturn mass planets \citep{Kanagawa15, Dong15, Dipierro15, Picogna15, Jin16}. 
If these gaps are indeed opened by planets, 
our estimated gas temperature and the derived depths of the HCO$^+$ gaps lead to planet masses of $\sim$0.8 $M_J$ at $\sim$28 AU and $\sim$2.1 $M_J$ at $\sim$69 AU, following the gap formula \citep{Kanagawa15} and assuming the viscosity $\alpha$ of 0.001. 
These values are larger than those published in the previous works based on the depths of the dust gaps and the dust temperature estimated from the continuum emission \citep{Kanagawa15, Dong15, Dipierro15, Jin16}. 
Gap opening by a planet is currently still unknown for a disk with a vertical temperature gradient (\citet{Lee15} and references therein). 
The gap formula and all the simulations simply assume vertically isothermal disks, 
but observations suggest cold dust near the disk mid-plane and hot gas above it \citep{Kwon11}. 
Nonetheless, sub-Jovian planetary masses can be obtained when the lower temperatures from these works are adopted in our estimation. 

Besides the gap depths, the noticeable offset between the D5 and D6 gaps and the outer HCO$^+$ gap may also evidence the presence of a planet. 
The B5 ring and the outer HCO$^+$ gap are almost located at the same radius, implying the presence of mm-sized dust at this radius. 
The coincidence of the HCO$^+$ gap and B5 ring is explained if the B5 ring corresponds to the horseshoe region of the putative planet \citep{Dipierro15, Picogna15}. 
Hence, the D5 and D6 gaps are offset from the horseshoe region, rather than aligning with the center of the outer HCO$^+$ gap.
In addition, our estimated gas temperature and HCO$^+$ column density may hint at a pressure bump at the outer edge of the outer HCO$^+$ gap, which situates around the bright continuum ring at a radius of $\sim$81.3 AU (B6 ring; \citet{ALMA15}). 
Such a pressure bump is normally associated with the outer edge of a planetary gap. 
In contrast, the relative position between the inner HCO$^+$ gaps and the D2 gap is less clear due to the limited S/N ratio and the uncertainties in deriving the HCO$^+$ column density.

\begin{figure}
\epsscale{0.8}
\plotone{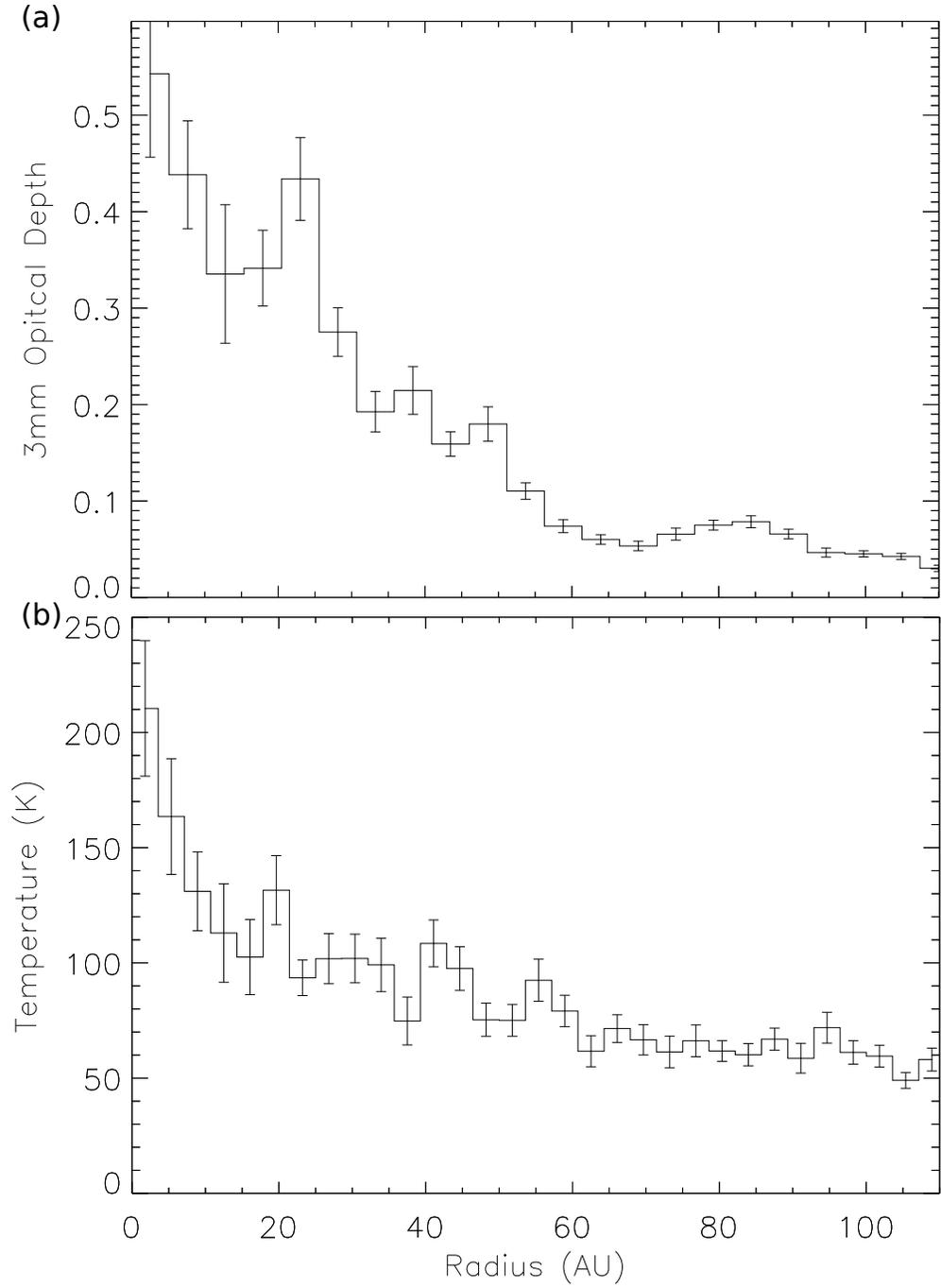}
\caption{(a) Derived radial profile of the optical depth of the 3 mm continuum emission in the HL Tau disk. (b) Estimated radial profile of the temperature of the HL Tau disk from the CO (1--0) image cube. Error bars correspond to their 1$\sigma$ uncertainties.}
\label{Tr}
\end{figure}

Alternatively, the origin of the outer HCO$^+$ gap can be related to the CO snow line without the presence of a planet. 
Although at a radius of $\gtrsim$60 AU, the temperature estimated from our CO image cube is $\sim$60 K (Fig.~\ref{Tr}), much higher than the CO sublimation temperature ($\sim$20 K), 
the CO depletion may still occur at the mid-plane, where the temperature can be as low as 20 K, as estimated from the millimeter continuum emission \citep{Kwon11,Pinte15}.
CO is a parent molecule of HCO$^+$, so its depletion can result in the depletion of HCO$^+$.
With the depletion considered, we are expecting to see that the HCO$^+$ column density decreases more rapidly beyond the CO snow line without forming a gap. 
Nevertheless, theoretical models suggest that gas can pile up around the snow line due to the transition in the strength of magneto-rotational instability, leading to a density bump \citep{Kretke07}.
With ALMA observations, changes in density and properties of dust around the CO snow line have been seen in the protoplanetary disk around HD 163296 \citep{Guidi16}.
With the combination of these two effects, the HCO$^+$ depletion and the density bump around the snow line, there is a possibility to form a gap in the HCO$^+$ column density profile.
Similar mechanisms have been proposed to explain the origin of the continuum gaps in the disk around HL Tau (e.g., \citet{Zhang15}).
However, to examine this scenario, detailed theoretical calculations on the required degrees of the HCO$^+$ depletion and the density bump to reproduce the observed outer HCO$^+$ gap are needed.
On the contrary, the temperature at the radius of the inner HCO$^+$ gap is most likely much higher than 20 K even at the mid-plane, as estimated from the millimeter continuum emission \citep{Kwon11,Pinte15}. 
Thus, this scenario is less possible to interpret the presence of the inner HCO$^+$ gap.

There are also other possibilities to create both gas and dust gaps in the disk around HL Tau in the absence of multiple planets, such as the variation of dust properties and dust sizes via dust-to-gas feedback (e.g., \citet{Takahashi14, Gonzalez15}). 
Our observations suggest that the total width of the D5 and D6 gaps (23.2 AU; \citet{Pinte15}) are approximately twice wider than that of the outer HCO$^+$ gap, 
while the D2 gap has a comparable width (11.0 AU; \citet{Pinte15}) to the inner HCO$^+$ gap. 
These hint at different gas-dust coupling along the disk radius \citep{Paardekooper04, Fouchet07}. 
Therefore, more detailed predications of locations, widths, and depths of dust and gas gaps from these models taking account of radial variation of gas-dust coupling is needed to examine these possibilities.
If planet-disk tidal interactions are indeed happening, the inferred epoch of formation of gas giant planets will occur much earlier than the predictions of the standard core accretion model \citep{Pollack96}, therefore calling for modified/alternative formation scenarios \citep{Piso15, Helled14, Chabrier14}.

\acknowledgements
This paper makes use of the following ALMA data: ADS/JAO.ALMA\#2011.0.00015.SV. ALMA is a partnership of ESO (representing its member states), NSF (USA) and NINS (Japan), together with NRC (Canada) and NSC and ASIAA (Taiwan), in cooperation with the Republic of Chile. The Joint ALMA Observatory is operated by ESO, AUI/NRAO and NAOJ.
We acknowledge I-Hsiu Lee for making an initial assessment of data quality.
N.H. acknowledges a grant from the Ministry of Science and Technology (MOST) of Taiwan (MOST 104-2119-M-001-016) in support of this work.
C.L. acknowledges a grant from MOST 104-2119-M-001-015-MY3 in support of this work.
S.T. acknowledges a grant from MOST 102-2119-M-001-012-MY3 in support of this work.

\clearpage

\begin{table}
\begin{center}
\caption[]{Summary of the HCO$^+$ Gaps}
\label{gap}
\begin{tabular}{ccccc}
\hline \hline \noalign {\smallskip}
Power-law Profile & Dip Column Density & Radius & FWHM Width  & Density Contrast \\
(10$^{15}$ cm$^{-2}$) & (10$^{15}$ cm$^{-2}$)  & (AU)  & (AU) \\
\hline \noalign {\smallskip}
$(16\pm2)\cdot(R/{\rm 20\ AU})^{-0.5\pm0.2}$ & 5$\pm$1 & 28$\pm$2 & 14$\pm$7 & 2.4$\pm$0.5 \\
$(9\pm1)\cdot(R/{\rm 60\ AU})^{-0.9\pm0.3}$ & 1.4$\pm$0.4 & 69$\pm$2 & 14$\pm$7 & 5.0$\pm$1.2 \\
\hline \hline \noalign {\smallskip}
\end{tabular}
\end{center}
\end{table}

\end{document}